\documentclass[twocolumn,showpacs,preprintnumbers,amsmath,amssymb,superscriptaddress,unsortedaddress]{revtex4}

\usepackage{graphicx}
\usepackage{dcolumn}
\usepackage{bm}

\begin{document}

\preprint{APS/123-QED}

\title{Incoherent neutral pion photoproduction on $^{12}$C}

\author{C.M.~Tarbert}
\author{D.P.~Watts}%
 \email{dwatts1@ph.ed.ac.uk}
\affiliation{School of Physics, University of Edinburgh, Edinburgh, UK}

\author{P.~Aguar}
\affiliation{Institut f\"ur Kernphysik, University of Mainz, Germany}

\author{J.~Ahrens}
\affiliation{Institut f\"ur Kernphysik, University of Mainz, Germany}

\author{J.R.M.~Annand} 
\affiliation{Department of Physics and Astronomy, University of Glasgow, Glasgow, UK}

\author{H.J.~Arends}
\affiliation{Institut f\"ur Kernphysik, University of Mainz, Germany}

\author{R.~Beck}
\affiliation{Institut f\"ur Kernphysik, University of Mainz, Germany}
\affiliation{Helmholtz-Institut f\"ur Strahlen- und Kernphysik, University Bonn, Germany}

\author{V. ~Bekrenev}
\affiliation{Petersburg Nuclear Physics Institute, Gatchina, Russia}

\author{B.~Boillat} 
\affiliation{Institut f\"ur Physik, University of Basel, Basel, Ch}

\author{A.~Braghieri} 
\affiliation{INFN Sezione di Pavia, Pavia, Italy}

\author{D.~Branford}
\affiliation{School of Physics, University of Edinburgh, Edinburgh, UK}

\author{W.J.~Briscoe}
\affiliation{Center for Nuclear Studies, The George Washington University, Washington, DC, USA}

\author{J.~Brudvik}
\affiliation{University of California at Los Angeles, Los Angeles, CA, USA}

\author{S.~Cherepnya}
\affiliation{ Lebedev Physical Institute, Moscow, Russia}

\author{R.~Codling} 
\affiliation{Department of Physics and Astronomy, University of Glasgow, Glasgow, UK}

\author{E.J.~Downie} 
\affiliation{Department of Physics and Astronomy, University of Glasgow, Glasgow, UK}

\author{K.~Foehl}
\affiliation{School of Physics, University of Edinburgh, Edinburgh, UK}

\author{D.I.~Glazier}
\affiliation{School of Physics, University of Edinburgh, Edinburgh, UK}

\author{P.~Grabmayr}
\affiliation{Physikalisches Institut Universit\"at T\"ubingen, T\"ubingen, Germany}

\author{R.~Gregor} 
\affiliation{ II. Physikalisches Institut,  University of Giessen, Germany}

\author{E.~Heid}
\affiliation{Institut f\"ur Kernphysik, University of Mainz, Germany}

\author{D.~Hornidge}
\affiliation{Mount Allison University, Sackville, NB, Canada} 

\author{O.~Jahn}
\affiliation{Institut f\"ur Kernphysik, University of Mainz, Germany}

\author{V.L.~Kashevarov} 
\affiliation{ Lebedev Physical Institute, Moscow, Russia}

\author{A.~Knezevic}
\affiliation{Rudjer Boskovic Institute, Zagreb, Croatia}

\author{R.~Kondratiev}  
\affiliation{Institute for Nuclear Research, Moscow, Russia}

\author{M.~Korolija}
\affiliation{Rudjer Boskovic Institute, Zagreb, Croatia}

\author{M.~Kotulla}
\affiliation{Institut f\"ur Physik, University of Basel, Basel, Ch}

\author{D.~Krambrich}
\affiliation{Institut f\"ur Kernphysik, University of Mainz, Germany}
\affiliation{Helmholtz-Institut f\"ur Strahlen- und Kernphysik, University Bonn, Germany}

\author{B.~Krusche}
\affiliation{Institut f\"ur Physik, University of Basel, Basel, Ch}

\author{M.~Lang}
\affiliation{Institut f\"ur Kernphysik, University of Mainz, Germany}
\affiliation{Helmholtz-Institut f\"ur Strahlen- und Kernphysik, University Bonn, Germany}

\author{V.~Lisin}
\affiliation{Institute for Nuclear Research, Moscow, Russia}

\author{K.~Livingston} 
\affiliation{Department of Physics and Astronomy, University of Glasgow, Glasgow, UK}

\author{S.~Lugert}
\affiliation{ II. Physikalisches Institut,  University of Giessen, Germany}

\author{I.J.D.~MacGregor} 
\affiliation{Department of Physics and Astronomy, University of Glasgow, Glasgow, UK}

\author{D.M.~Manley}
\affiliation{Kent State University, Kent, OH, USA}

\author{M.~Martinez}
\affiliation{Institut f\"ur Kernphysik, University of Mainz, Germany}

\author{J.C.~McGeorge} 
\affiliation{Department of Physics and Astronomy, University of Glasgow, Glasgow, UK}

\author{D.~Mekterovic}
\affiliation{Rudjer Boskovic Institute, Zagreb, Croatia}

\author{V.~Metag}
\affiliation{ II. Physikalisches Institut,  University of Giessen, Germany}

\author{B.M.K.~Nefkens} 
\affiliation{University of California at Los Angeles, Los Angeles, CA, USA}

\author{A.~Nikolaev}
\affiliation{Institut f\"ur Kernphysik, University of Mainz, Germany}
\affiliation{Helmholtz-Institut f\"ur Strahlen- und Kernphysik, University Bonn, Germany}

\author{R.~Novotny}
\affiliation{ II. Physikalisches Institut,  University of Giessen, Germany}

\author{R.O.~Owens}
\affiliation{Department of Physics and Astronomy, University of Glasgow, Glasgow, UK}

\author{P.~Pedroni}
\affiliation{INFN Sezione di Pavia, Pavia, Italy}

\author{A.~Polonski}
\affiliation{Institute for Nuclear Research, Moscow, Russia}

\author{S.N.~Prakhov}
\affiliation{University of California at Los Angeles, Los Angeles, CA, USA}

\author{J.W.~Price}
\affiliation{University of California at Los Angeles, Los Angeles, CA, USA}

\author{G.~Rosner}
\affiliation{Department of Physics and Astronomy, University of Glasgow, Glasgow, UK}

\author{M.~Rost}
\affiliation{Institut f\"ur Kernphysik, University of Mainz, Germany}

\author{T.~Rostomyan}
\affiliation{INFN Sezione di Pavia, Pavia, Italy}

\author{S.~Schadmand}
\affiliation{ II. Physikalisches Institut,  University of Giessen, Germany}

\author{S.~Schumann}
\affiliation{Institut f\"ur Kernphysik, University of Mainz, Germany}
\affiliation{Helmholtz-Institut f\"ur Strahlen- und Kernphysik, University Bonn, Germany}

\author{D.~Sober}
\affiliation{The Catholic University of America, Washington, DC, USA}

\author{A.~Starostin}
\affiliation{University of California at Los Angeles, Los Angeles, CA, USA}

\author{I.~Supek}
\affiliation{Rudjer Boskovic Institute, Zagreb, Croatia}

\author{A.~Thomas}
\affiliation{Institut f\"ur Kernphysik, University of Mainz, Germany}

\author{M.~Unverzagt}
\affiliation{Institut f\"ur Kernphysik, University of Mainz, Germany}

\author{Th.~Walcher}
\affiliation{Institut f\"ur Kernphysik, University of Mainz, Germany}

\author{F.~Zehr}
\affiliation{Institut f\"ur Physik, University of Basel, Basel, Ch}


\collaboration{The Crystal Ball at MAMI and A2 Collaboration}
%


\date{\today}

\begin{abstract}
We present the first detailed measurement of incoherent photoproduction of neutral pions to a discrete state of a residual nucleus. The $^{12}$C$(\gamma,\pi^{0})^{12}$C$^{*}_{4.4MeV}$ reaction has been studied with the Glasgow photon tagger at MAMI employing a new technique which uses the large solid angle Crystal Ball detector both as a $\pi^{0}$ spectrometer and to detect decay photons from the excited residual nucleus. The technique has potential applications to a broad range of future nuclear measurements with the Crystal Ball and similar detector systems elsewhere. The data are sensitive to the propagation of the $\Delta$ in the nuclear medium and will give the first information on matter transition form factors from measurements with an electromagnetic probe. The incoherent cross sections are compared to two theoretical predictions including a $\Delta$-hole model.

\end{abstract}

\maketitle


This Letter reports the first detailed measurement of nuclear $\pi^{0}$ photoproduction populating a specific excited state in the residual nucleus. The  photoproduction of $\pi^{0}$s from nuclei at intermediate photon energies is of great interest for a number of reasons. The dominance of the $\Delta$ resonance in the $\pi^{0}$ photoproduction amplitude and the ability of the electromagnetic probe to sample the full nuclear volume makes the reaction one of the cleanest tests of our understanding of the interaction of the $\Delta$ in the nuclear environment. This dominance of the $\Delta$ in the production amplitude has a further useful consequence in that it leads to an approximately equal probability for $\pi^{0}$ photoproduction from both protons and neutrons in the nucleus. Potentially this allows access to accurate information on the transition form factor for reactions in which the dominant change takes place in the neutron wave function, while circumventing many of the difficulties present in traditional methods using strongly interacting probes. Measurement of this incoherent process to discrete nuclear states also offers opportunities to use the spin-isospin selection rules to study specific components of the basic pion photoproduction amplitude.

 The importance of the incoherent $(\gamma,\pi^{0})$ process from nuclear targets has been appreciated for some time \cite{Takaki:1985vm,trya:1990,trya:2007}. However, although nuclear $\pi^{0}$ photoproduction has been studied at various facilities for over 30 years, no results for the population of discrete residual nuclear states have been obtained because the accuracy of the angle and the energy determination of the photons from the $\pi^{0}\rightarrow 2\gamma$ decay needed to resolve states in the residual nucleus has not been achieved. The only published information on the incoherent $(\gamma,\pi^{0})$ reaction was extracted from measurements with an untagged photon beam  \cite{decg:00} which obtained the integrated yield of decay photons from the 3.56-MeV state in $^{6}$Li and the 4.4-MeV state of $^{12}$C. The 3.56-MeV, 0$^{+}$, T=1 state in $^{6}$Li is reached only via the weak isoscalar single nucleon amplitude and a low cross section is observed illustrating the value of the incoherent $\pi^{0}$ reaction for isolating the smaller amplitude terms. An unpublished study of the  4.4-MeV yield in a restricted $\pi^{0}$ angle range obtained at MAMI is presented in Ref. \cite{Schmitz:1996}. There is also some information on the summed ``non-coherent'' strength, presented in Refs  \cite{Koch:1989xf,Arends:1982cw} for $^{12}$C and $^{40}$Ca. However, it is difficult to extract detailed information from the ``non-coherent'' strength as it includes both the incoherent processes and the quasi-free process in which nucleons are also ejected. The general features of these data were described by a Fermi gas model of the quasi-free process \cite{Arends:1982cw}.

 There are at present only two available calculations of incoherent $\pi^{0}$ photoproduction to discrete residual nuclear states, both for the $^{12}$C nucleus. The most detailed treatment \cite{Takaki:1985vm} is based on the $\Delta$-hole model  and includes a study of the contributions of various $\pi^{0}$ and $\Delta$-nucleus interaction processes to the incoherent cross section. The predictions highlight the sensitivity of the incoherent reaction to the character of the nuclear transition involved and to specific $\Delta$-nucleus processes such as  $\Delta N$ interactions, which have a much smaller effect on other observables such as the coherent cross section. The other calculation \cite{trya:2007} is less sophisticated. It uses the plane wave impulse approximation and makes a rough estimate of the effect of the $\pi^{0}$-nucleus final-state interaction. Very importantly, however, this treatment does derive formulae for the angular correlation between the emitted $\pi^{0}$ and the subsequent nuclear decay photon. This correlation turns out to be strong and its use is essential in this data analysis. Theoretical work, now in progress \cite{tiator:priv}, will give additional predictions of the incoherent cross section based on the Mainz unitary isobar model \cite{Maid:1999} with a complex pion optical potential and $\Delta$ medium modifications incorporated using a $\Delta$ self-energy.



\begin{figure}
\includegraphics[scale=0.8]{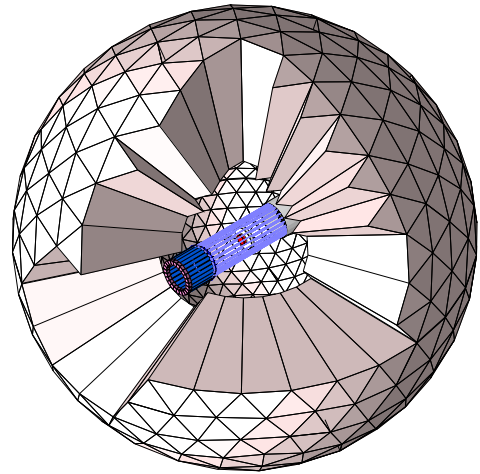}
\caption{\label{fig:cb} Diagram showing the Crystal Ball detector, the $^{12}$C target (red) and the the PID detector (blue). The MWPC is omitted for clarity}
\end{figure}

The data presented here were obtained as part of a series of experiments on neutral pion photoproduction from $^{12}$C, $^{16}$O, $^{40}$Ca and $^{208}$Pb targets, carried out with the Crystal Ball (CB) detector \cite{PhysRevD.25.2259} and the Glasgow photon tagger \cite{Hall:1996gh,Anthony:1991eq} at MAMI \cite{Herminghaus:1983nv}. The CB (Figure \ref{fig:cb}) is a 672 element NaI detector covering ~94\% of 4$\pi$. Photons incident on the ball produce an electromagnetic shower that typically deposits 98\% of its energy in a cluster of 13 crystals. Analysis of the center of gravity of the shower allows angular resolutions for the photon of ~2-3$^{\circ}$. The high light output of NaI also permits a good determination of the photon energy ( $\frac{\sigma}{E_{\gamma}} \sim \frac{1.7}{E_{\gamma}}(GeV)^{0.4}$). Since its move to Mainz there has been a complete overhaul of the electronics for the CB \cite{DirkKram}  and it has been instrumented with additional detectors. A central detector providing charged particle identification \cite{Watts:2004xb} was provided by the Edinburgh and Glasgow groups and two cylindrical Multi Wire Proportional Counters (MWPC) were transferred from Daphne \cite{Audit:1991gq}.  The forward hole of the CB was instrumented with the TAPS detector array \cite{Novotny:1991ht}, but this was not used in the present analysis.

The tagged photon beam covered the energy range 120 to 819 MeV with a tagging resolution of $\sim$2~MeV full width and an intensity of $\sim2\times10^{5}$~$\gamma$ s$^{-1}$~MeV$^{-1}$. The tagged photons were incident on a 1.5~cm thick $^{12}$C target. Emitted photons were detected in coincidence in the CB, with additional information on charged particles given by the central detectors. The reconstructed vertex position from multiple charged track events in the MWPC allows accurate reconstruction (to $\sim\frac{1}{2}$mm) of the target position relative to the CB.


\begin{figure}
\includegraphics[scale=0.4]{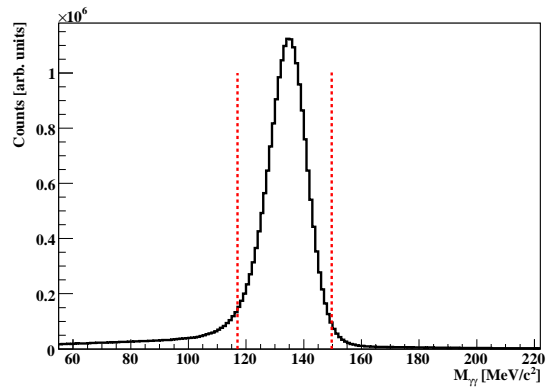}
\caption{\label{fig:imass} The spectrum of invariant mass reconstructed from the 2$\gamma$ events in the CB for $E_{\gamma}\leq$ 400~MeV. Events in the mass range 117-149 MeV were selected for the analysis.
}
\end{figure}

Neutral pions were identified in the CB from their 2$\gamma$ decay. The invariant mass spectrum reconstructed from the detected 2$\gamma$ events in the CB is presented in Fig. \ref{fig:imass}. The contribution of pions not originating from the $^{12}$C target was found to be only $\sim$3\% in additional runs with the target removed. 
  
\begin{figure}
\rotatebox{0}{\includegraphics[scale=0.4]{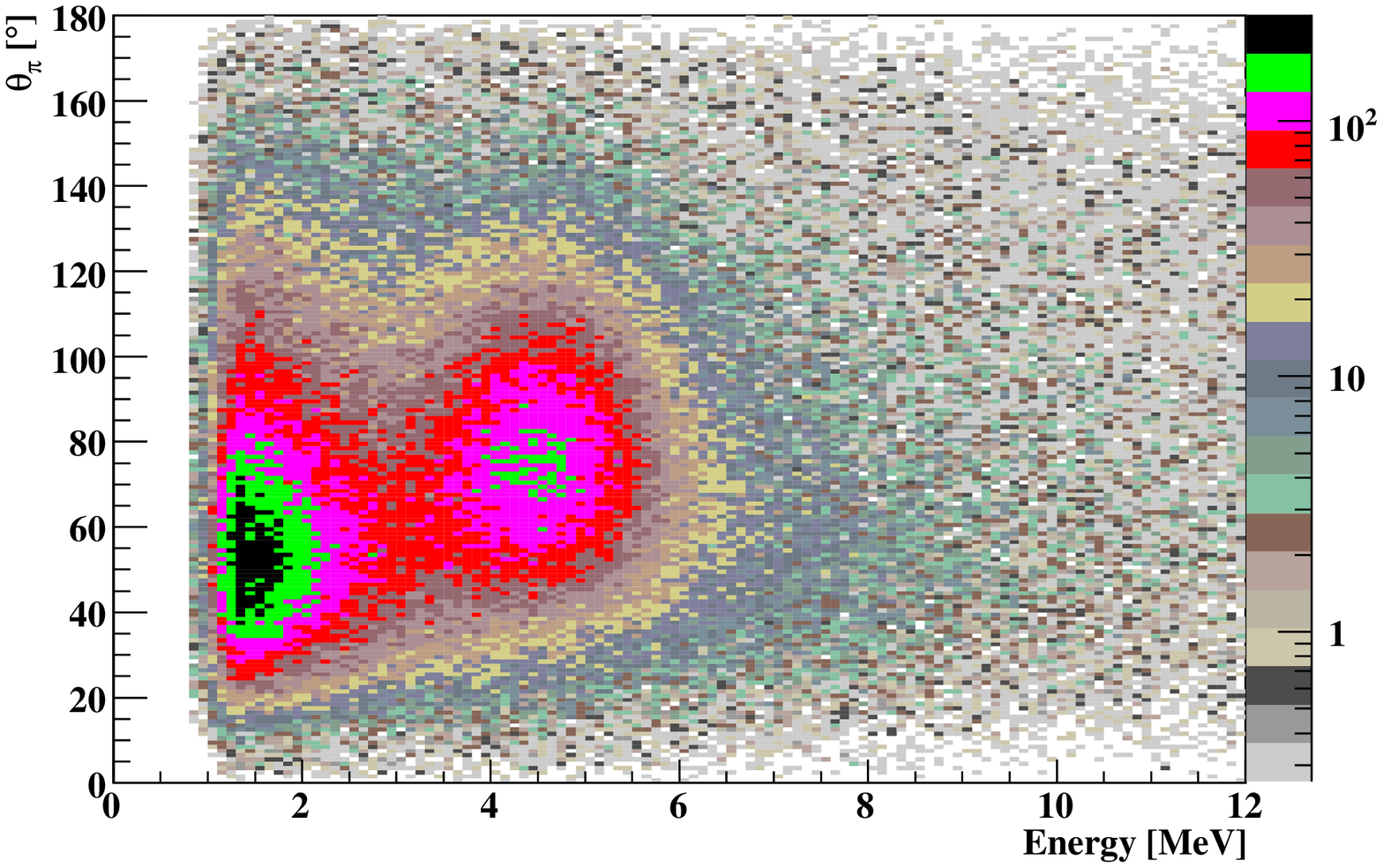}}
\rotatebox{0}{\includegraphics[scale=0.4]{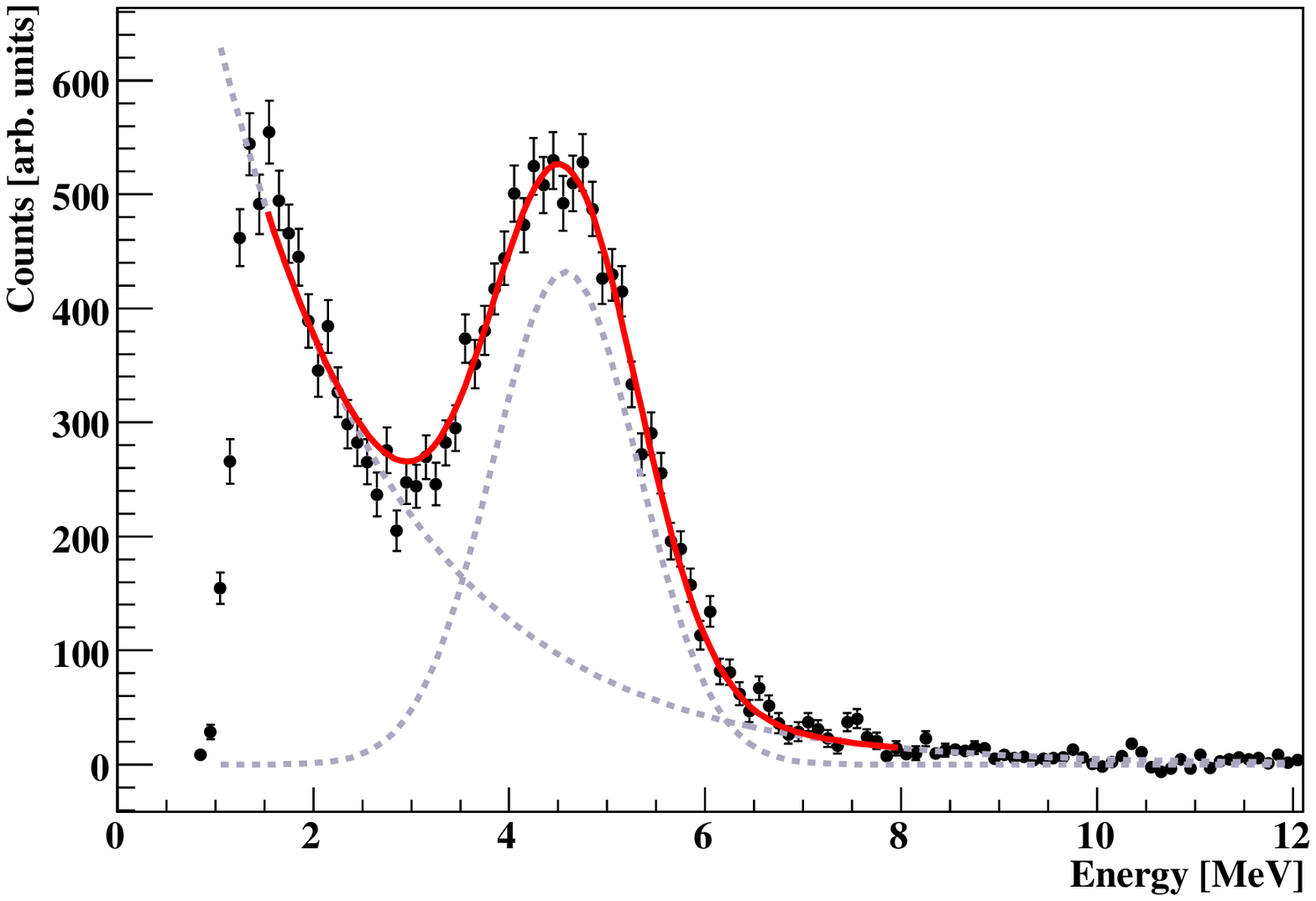}}
\caption{\label{fig:c12dg} {\bf Upper plot:} $\theta_{\pi}$ versus the energy of the low-energy clusters detected in the CB. {\bf Lower plot:} projection of the energy distribution for $\theta_{\pi}$ bin of 78$\pm$2$^{\circ}$. The gray dotted lines show the result of an  exponential plus Gaussian fit to the data. Both plots for $E_{\gamma}$=235$\pm$10 MeV and for  $E_{\pi}^{diff}$ below 20 MeV.}
\end{figure}

The energy difference ($E_{\pi}^{diff}$) between the reconstructed $\pi^{0}$ energy and its calculated energy (using the tagged photon energy, measured $\theta_{\pi}$ and assuming coherent $\pi^{0}$ photoproduction) was restricted to less than 20 MeV to suppress the contribution of quasi-free $\pi^{0}$ production. 
Figure \ref{fig:c12dg} (upper) shows a plot of the $\pi^{0}$ polar angle versus the energy of the time-correlated low energy photon clusters in the CB for these data. 
Figure 3 (lower) shows the projection of the photon energy distribution for the angular range $\theta_{\pi}$=78$\pm$2$^{\circ}$. Nuclear decay photons from the 4.4-MeV state in $^{12}$C are clear in both plots. There is no evidence of significant nuclear decay radiation from higher-lying residual states. A smoothly falling background of low-energy photons is also evident, whose distribution of strength with $\pi^{0}$ angle appears strongly correlated with the coherent cross section (which is maximum at $\theta_{\pi}$$\sim$50$^{\circ}$ \cite{Ruth,Claire} for the chosen incident $E_{\gamma}$ bin). GEANT3 (G3) simulations (not shown) confirm the dominant cause of this background to be low energy photons, which split off from the $\pi^{0}$ decay photon clusters. A reduction in the contribution of split off photons is achieved in the present analysis by requiring that low energy photons have angular separation of greater than $35^{\circ}$ from either of the $\pi^{\circ}$ decay photons.


\begin{figure}
\rotatebox{0}{\includegraphics[scale=0.45]{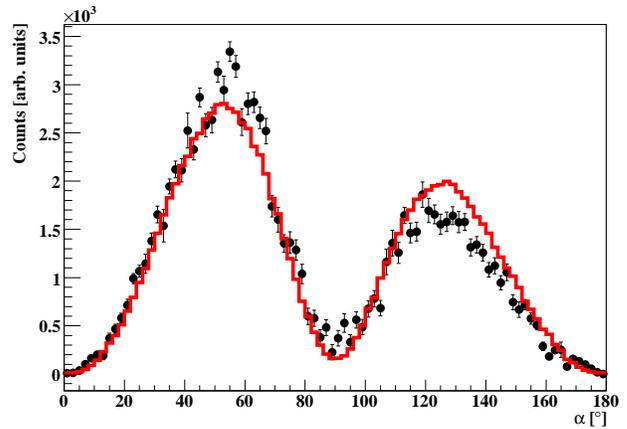}}
\caption{\label{fig:wdist}Comparison of the distribution of the $\alpha$, the polar angle of the nuclear decay photon with respect to the recoil direction, for incident $E_{\gamma}$=235$\pm$10 MeV (black points) with the distribution suggested in Ref. \cite{trya:2007}, passed through the G3 simulation of the experimental acceptance (red line). 
}
\end{figure}

\begin{figure}
\includegraphics[scale=0.4]{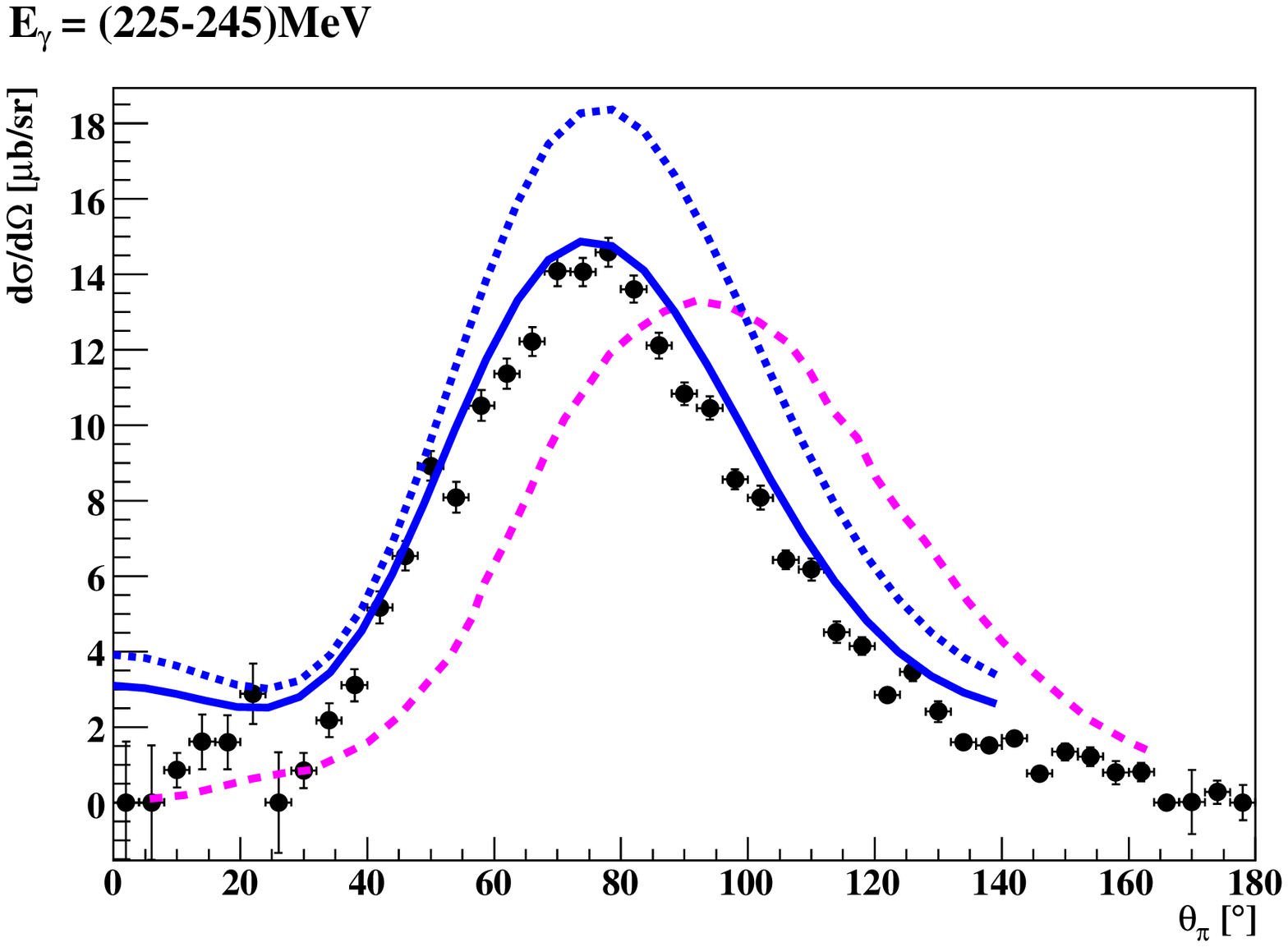}
\includegraphics[scale=0.4]{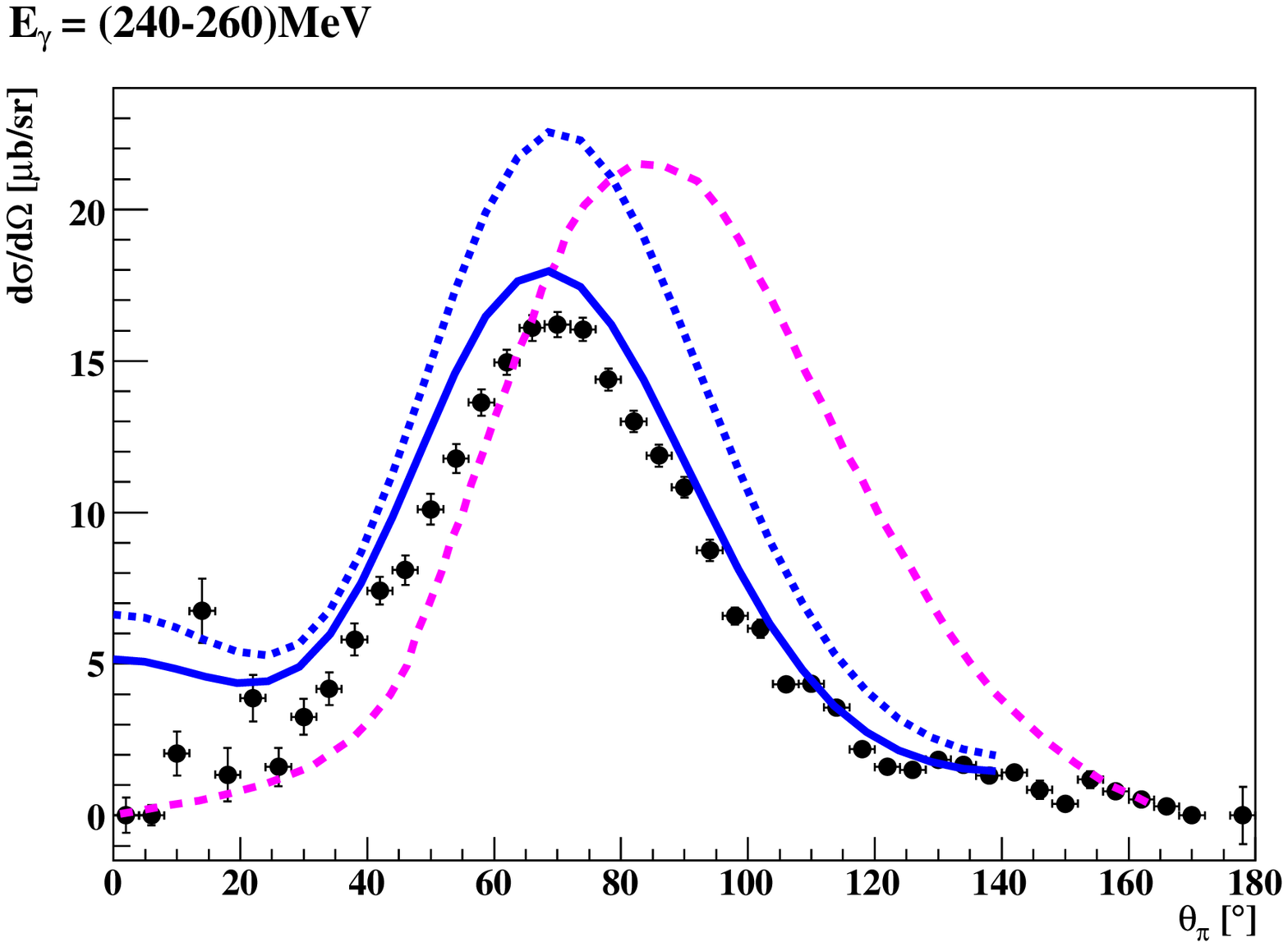}
\includegraphics[scale=0.4]{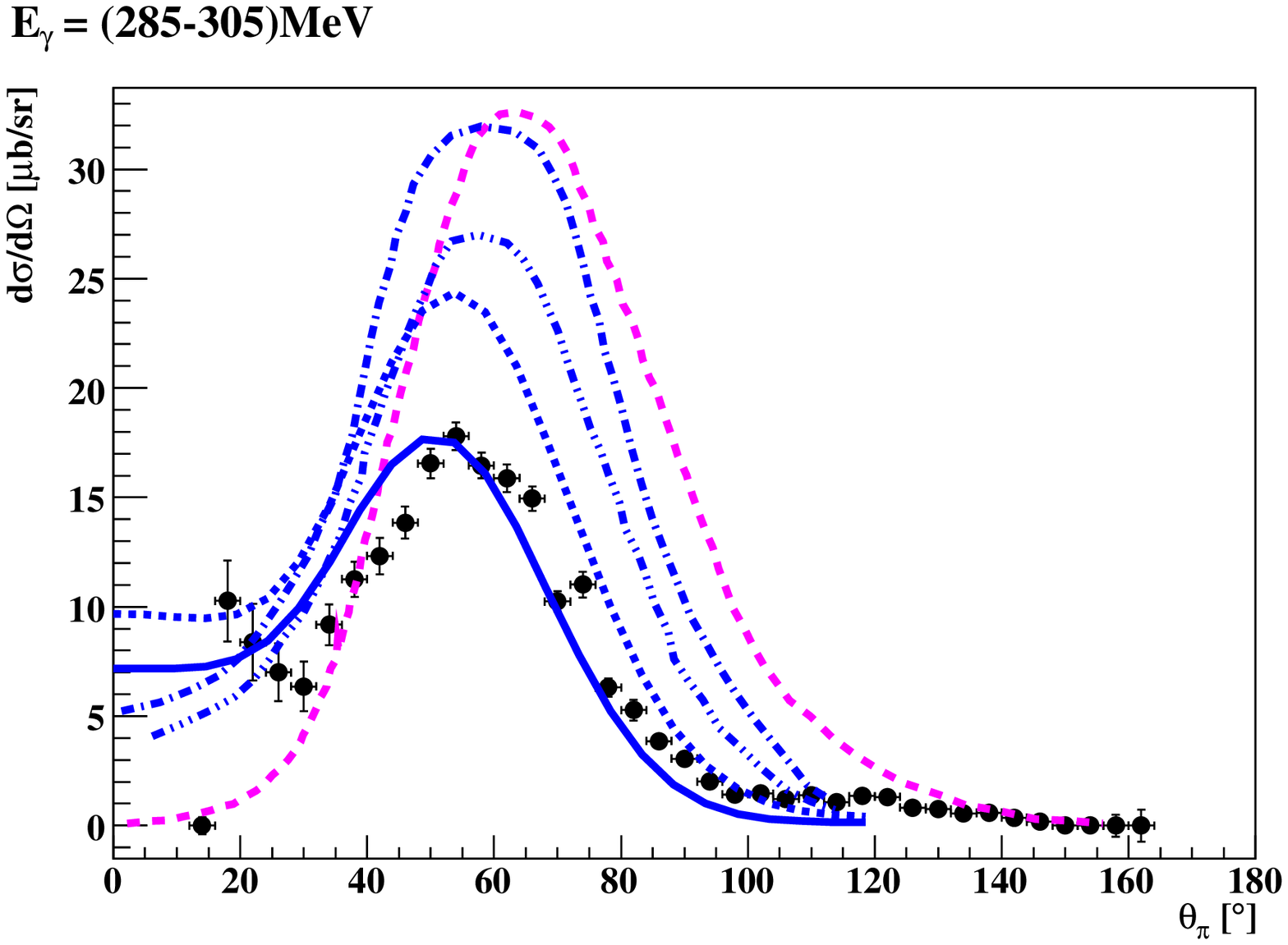}

\caption{\label{fig:angdist}The $^{12}$C$(\gamma,\pi^{0})$$^{12}$C$_{4.4 MeV}$ cross section presented as a function of $\theta_{\pi}$ for $E_{\gamma}$ bins indicated in the figure. The predictions by Takaki et al. \cite{Takaki:1985vm,takaki:priv} are shown by blue lines: modified DWIA (dot-dash); many body effects added (dot-dot-dash); multistep mechanisms also added (dash); $\Delta$-N interaction also added (solid). Pink long dashed line shows predictions of Ref. \cite{trya:2007}}
\end{figure}

To extract the incoherent cross section to the 4.4-MeV state, the low-energy photon spectrum for each $\theta_{\pi}$ bin was fitted with a Gaussian centered at 4.4~MeV plus an exponential background. The fitted components are shown in Fig. \ref{fig:c12dg} (lower). The shape of the background was consistent with the energy distribution predicted by G3. The fitted background also accounts for the small fraction ($\sim$4\%) of random events in the $\sim$30ns wide $\pi^{0}$--$\gamma$ coincidence peak. In principle the strength of the observed 4.4-MeV peak may contain contributions from higher-lying states cascading through this state.  The contribution from such cascades was quantified from the ratio of double to single low-energy photon detection rates in the CB and found to be less than 5\% of the 4.4-MeV yield. This is expected as the strongest branching ratio to the 4.4-MeV state is 2.1\% for the 15.1~MeV ($1^{+},T=1$) state,  which is not produced via $\Delta$ excitation \cite{Takaki:1985vm}, and the $\gamma$ branch for other states is at least a factor of 10 smaller.

To convert the incoherent yield to the 4.4-MeV state at a particular pion angle into a cross section, the efficiency of the CB for simultaneous detection of both the $\pi^{0}$ and the 4.4~MeV decay $\gamma$ averaged over the angular distribution between them is required. This was obtained from the G3 simulation. The required $\pi^{0}$--$\gamma$ angular correlation was taken from Ref. \cite{trya:2007} where it is given in terms of the angle $\alpha$ between the decay photon and the $^{12}$C recoil direction which has the distribution $\frac{15}{8}\sin^{2}(2\alpha)$. The  combined $\pi^{0}$--$\gamma$ detection efficiency so obtained  varies from $\sim$20 to 30\% over the $\pi^{0}$ angular range 30-160$^{\circ}$ but is smaller outside this range due to holes in the CB at forward and backward angles. It was used to extract the differential $\pi^{0}$ production cross sections for the 4.4-MeV state shown in fig. \ref{fig:angdist}. 

The validity of the  $\pi^{0}$--$\gamma$ angular correlation obtained in Ref. \cite{trya:2007}, which was used to calculate the overall CB efficiency, was checked by plotting the experimental distribution of the angle $\alpha$. Figure \ref{fig:wdist} shows the data for incident energies $E_{\gamma}$=235$\pm$10~MeV. This is compared with a G3 simulation, which uses as its input the predicted $\alpha$-distribution,  $\frac{15}{8}\sin^{2}(2\alpha)$, and a $\pi^{0}$ angular distribution which has the same shape as the data shown in fig. \ref{fig:angdist}. The agreement between the data and the simulated $\alpha$-distribution of the decay photons clearly establishes the dominance of the $\sin^{2}(2\alpha)$ term in their angular distribution. The polarization state of the recoil $^{12}$C nuclei, which leads to the simple predicted distribution shape, results from the dominance of the spin independent terms in the $(\gamma,\pi^{0})$ amplitude on a single nucleon. In fact the calculation of Ref. \cite{trya:2007} suggests that the spin-dependent terms will also provide a contribution to the incoherent excitation of the 4.4-MeV state at the few percent level and that this contribution will have a cos$^{2}(2\alpha)$ distribution. Such a contribution may account for some of the remaining discrepancy between the data and the prediction in Fig. \ref{fig:wdist}. It is clear that angular correlation data of this type will be valuable in separating the components of the basic photoproduction amplitude.

Differential cross sections for incoherent $\pi^{0}$ photoproduction from $^{12}$C populating the $2^{+}$ state at 4.4 MeV are shown in Fig. \ref{fig:angdist} where they are compared with the two available calculations. The main sources of systematic uncertainty in the present measurement arise from the the detection efficiency calculations and the yield extraction technique with smaller uncertainties arising from the photon flux determination and the measurement of the target thickness. The total systematic uncertainty in the cross sections is estimated to be $\sim\pm$10\%.

The shape of the angular distributions in Fig. \ref{fig:angdist} is determined basically by the momentum dependence of the transition form factor between the 4.4-MeV state and the ground state, although pion distortion is expected \cite{Takaki:1985vm} to enhance the cross section at small angles and shift the main peak to larger angles by a few degrees. The calculation of Ref. \cite{Takaki:1985vm} uses nuclear wave functions obtained by fitting elastic and inelastic electron scattering form factors and includes as full a treatment of $\pi^{0}$-- and $\Delta$ -- interactions in the nucleus as can be achieved in the $\Delta$-hole model. It is reassuring, therefore, to see that the angular distribution shape is very well described. The magnitude of the theoretical cross sections is mainly affected by the details of the pion and $\Delta$ interactions in the nucleus. The four $E_{\gamma}$=295~MeV curves from Ref \cite{Takaki:1985vm} in Fig. \ref{fig:angdist} chart the reduction in the cross section as many body effects, multistep mechanisms and the $\Delta$-N interaction are successively introduced. Given the large combined change produced by these factors, the full calculation gives a fairly good explanation of the results. Additional experimental data covering a wider photon energy range would probably help identify which parts of the calculation are not yet adequate. The calculation of Ref. \cite{trya:2007}, which is basically a plane-wave treatment and uses wave functions from an L-S coupling model does significantly less well in explaining the shape, magnitude and photon energy dependence of the measured cross sections.

In summary, the present experiment is the first detailed measurement of incoherent $\pi^{0}$ photoproduction from a nucleus and employs a novel nuclear decay photon technique that will have application to further nuclear measurements at the CB and other experimental facilities. The incoherent cross sections are in general agreement with the available $\Delta$-hole model calculation, but the comparison indicates refinements in the calculation may be necessary. The extracted incoherent cross sections will also be important in improving the suppression of incoherent background in the extraction of the coherent $\pi^{0}$ production process \cite{Krusche:2002iq,Watts:Cohpi}, the poor determination of which has previously limited attempts to obtain accurate measurements of the matter form factors of nuclei \cite{Krusche:2005jx}.

\begin{acknowledgments}
The authors wish to thank T. Takaki for calculational results and valuable comments and to acknowledge the excellent support of
the accelerator group of MAMI. This work was supported
by the UK EPSRC, the Deutsche Forschungsgemeinschaft (SFB
443, SFB/Transregio16 and the European Community-Research Infrastructure Activity under
the FP6 "Structuring the European Research Area" programme
(HadronPhysics, contract number RII3-CT-2004-506078), the USDOE, USNSF and NSERC (Canada). We thank the undergraduates students of Mount Allison and George Washington
Universities for their assistance.

\end{acknowledgments}





\bibliography{incohpi}

\begin{thebibliography}{23}
\expandafter\ifx\csname natexlab\endcsname\relax\def\natexlab#1{#1}\fi
\expandafter\ifx\csname bibnamefont\endcsname\relax
  \def\bibnamefont#1{#1}\fi
\expandafter\ifx\csname bibfnamefont\endcsname\relax
  \def\bibfnamefont#1{#1}\fi
\expandafter\ifx\csname citenamefont\endcsname\relax
  \def\citenamefont#1{#1}\fi
\expandafter\ifx\csname url\endcsname\relax
  \def\url#1{\texttt{#1}}\fi
\expandafter\ifx\csname urlprefix\endcsname\relax\def\urlprefix{URL }\fi
\providecommand{\bibinfo}[2]{#2}
\providecommand{\eprint}[2][]{\url{#2}}

\bibitem[{\citenamefont{Takaki et~al.}(1985)\citenamefont{Takaki, Suzuki, and
  Koch}}]{Takaki:1985vm}
\bibinfo{author}{\bibfnamefont{T.}~\bibnamefont{Takaki}},
  \bibinfo{author}{\bibfnamefont{T.}~\bibnamefont{Suzuki}}, \bibnamefont{and}
  \bibinfo{author}{\bibfnamefont{J.~H.} \bibnamefont{Koch}},
  \bibinfo{journal}{Nucl. Phys.} \textbf{\bibinfo{volume}{A443}},
  \bibinfo{pages}{570} (\bibinfo{year}{1985}).

\bibitem[{\citenamefont{Tryasuchev and Kolchin}(2007)}]{trya:2007}
\bibinfo{author}{\bibfnamefont{V.~A.} \bibnamefont{Tryasuchev}}
  \bibnamefont{and} \bibinfo{author}{\bibfnamefont{A.~V.}
  \bibnamefont{Kolchin}}, \bibinfo{journal}{Yad. Phys.}
  \textbf{\bibinfo{volume}{70}}, \bibinfo{pages}{861} (\bibinfo{year}{2007}).

\bibitem[{\citenamefont{Tryasuchev and Kolchin}(1991)}]{trya:1990}
\bibinfo{author}{\bibfnamefont{V.~A.} \bibnamefont{Tryasuchev}}
  \bibnamefont{and} \bibinfo{author}{\bibfnamefont{A.~V.}
  \bibnamefont{Kolchin}}, \bibinfo{journal}{Yad. Phys.}
  \textbf{\bibinfo{volume}{53}}, \bibinfo{pages}{703} (\bibinfo{year}{1991}).

\bibitem[{\citenamefont{Naumenko et~al.}(1989)\citenamefont{Naumenko, Repenko,
  Stibunov, and Tryasuchev}}]{decg:00}
\bibinfo{author}{\bibfnamefont{G.~A.} \bibnamefont{Naumenko}},
  \bibinfo{author}{\bibfnamefont{E.~V.} \bibnamefont{Repenko}},
  \bibinfo{author}{\bibfnamefont{V.~N.} \bibnamefont{Stibunov}},
  \bibnamefont{and} \bibinfo{author}{\bibfnamefont{V.~A.}
  \bibnamefont{Tryasuchev}}, \bibinfo{journal}{JETP Lett.}
  \textbf{\bibinfo{volume}{50}}, \bibinfo{pages}{251} (\bibinfo{year}{1989}).

\bibitem[{\citenamefont{Schmitz}(1996)}]{Schmitz:1996}
\bibinfo{author}{\bibfnamefont{M.}~\bibnamefont{Schmitz}},
  \bibinfo{journal}{Ph.D. Thesis, Johannes Gutenberg-Universitat Mainz}
  (\bibinfo{year}{1996}).

\bibitem[{\citenamefont{Koch et~al.}(1989)}]{Koch:1989xf}
\bibinfo{author}{\bibfnamefont{G.}~\bibnamefont{Koch}} \bibnamefont{et~al.},
  \bibinfo{journal}{Phys. Rev. Lett.} \textbf{\bibinfo{volume}{63}},
  \bibinfo{pages}{498} (\bibinfo{year}{1989}).

\bibitem[{\citenamefont{Arends et~al.}(1983)}]{Arends:1982cw}
\bibinfo{author}{\bibfnamefont{J.}~\bibnamefont{Arends}} \bibnamefont{et~al.},
  \bibinfo{journal}{Z. Phys.} \textbf{\bibinfo{volume}{A311}},
  \bibinfo{pages}{367} (\bibinfo{year}{1983}).

\bibitem[{\citenamefont{Tiator and Kamalov}(2007)}]{tiator:priv}
\bibinfo{author}{\bibfnamefont{L.}~\bibnamefont{Tiator}} \bibnamefont{and}
  \bibinfo{author}{\bibfnamefont{S.}~\bibnamefont{Kamalov}},
  \bibinfo{journal}{Private communication}  (\bibinfo{year}{2007}).

\bibitem[{\citenamefont{Drechsel et~al.}(1999)\citenamefont{Drechsel, Hanstein,
  Kamalov, and Tiator}}]{Maid:1999}
\bibinfo{author}{\bibfnamefont{D.}~\bibnamefont{Drechsel}},
  \bibinfo{author}{\bibfnamefont{O.}~\bibnamefont{Hanstein}},
  \bibinfo{author}{\bibfnamefont{S.~S.} \bibnamefont{Kamalov}},
  \bibnamefont{and} \bibinfo{author}{\bibfnamefont{L.}~\bibnamefont{Tiator}},
  \bibinfo{journal}{Nucl. Phys.} \textbf{\bibinfo{volume}{A645}},
  \bibinfo{pages}{145} (\bibinfo{year}{1999}).

\bibitem[{\citenamefont{Oreglia et~al.}(1982)\citenamefont{Oreglia, Bloom,
  Bulos, Chestnut, Gaiser, Godfrey, Kiesling, Lockman, Scharre, Partridge
  et~al.}}]{PhysRevD.25.2259}
\bibinfo{author}{\bibfnamefont{M.}~\bibnamefont{Oreglia}},
  \bibinfo{author}{\bibfnamefont{E.}~\bibnamefont{Bloom}},
  \bibinfo{author}{\bibfnamefont{F.}~\bibnamefont{Bulos}},
  \bibinfo{author}{\bibfnamefont{R.}~\bibnamefont{Chestnut}},
  \bibinfo{author}{\bibfnamefont{J.}~\bibnamefont{Gaiser}},
  \bibinfo{author}{\bibfnamefont{G.}~\bibnamefont{Godfrey}},
  \bibinfo{author}{\bibfnamefont{C.}~\bibnamefont{Kiesling}},
  \bibinfo{author}{\bibfnamefont{W.}~\bibnamefont{Lockman}},
  \bibinfo{author}{\bibfnamefont{D.~L.} \bibnamefont{Scharre}},
  \bibinfo{author}{\bibfnamefont{R.}~\bibnamefont{Partridge}},
  \bibnamefont{et~al.}, \bibinfo{journal}{Phys. Rev. D}
  \textbf{\bibinfo{volume}{25}}, \bibinfo{pages}{2259} (\bibinfo{year}{1982}).

\bibitem[{\citenamefont{Hall et~al.}(1996)\citenamefont{Hall, Miller, Beck, and
  Jennewein}}]{Hall:1996gh}
\bibinfo{author}{\bibfnamefont{S.~J.} \bibnamefont{Hall}},
  \bibinfo{author}{\bibfnamefont{G.~J.} \bibnamefont{Miller}},
  \bibinfo{author}{\bibfnamefont{R.}~\bibnamefont{Beck}}, \bibnamefont{and}
  \bibinfo{author}{\bibfnamefont{P.}~\bibnamefont{Jennewein}},
  \bibinfo{journal}{Nucl. Instrum. Meth.} \textbf{\bibinfo{volume}{A368}},
  \bibinfo{pages}{698} (\bibinfo{year}{1996}).

\bibitem[{\citenamefont{Anthony et~al.}(1991)\citenamefont{Anthony, Kellie,
  Hall, Miller, and Ahrens}}]{Anthony:1991eq}
\bibinfo{author}{\bibfnamefont{I.}~\bibnamefont{Anthony}},
  \bibinfo{author}{\bibfnamefont{J.~D.} \bibnamefont{Kellie}},
  \bibinfo{author}{\bibfnamefont{S.~J.} \bibnamefont{Hall}},
  \bibinfo{author}{\bibfnamefont{G.~J.} \bibnamefont{Miller}},
  \bibnamefont{and} \bibinfo{author}{\bibfnamefont{J.}~\bibnamefont{Ahrens}},
  \bibinfo{journal}{Nucl. Instrum. Meth.} \textbf{\bibinfo{volume}{A301}},
  \bibinfo{pages}{230} (\bibinfo{year}{1991}).

\bibitem[{\citenamefont{Herminghaus et~al.}(1983)}]{Herminghaus:1983nv}
\bibinfo{author}{\bibfnamefont{H.}~\bibnamefont{Herminghaus}}
  \bibnamefont{et~al.}, \bibinfo{journal}{IEEE Trans. Nucl. Sci.}
  \textbf{\bibinfo{volume}{30}}, \bibinfo{pages}{3274} (\bibinfo{year}{1983}).

\bibitem[{\citenamefont{Krambrich}(2006)}]{DirkKram}
\bibinfo{author}{\bibfnamefont{D.}~\bibnamefont{Krambrich}},
  \bibinfo{journal}{Dissertation, Universitat Mainz}  (\bibinfo{year}{2006}).

\bibitem[{\citenamefont{Watts}(2004)}]{Watts:2004xb}
\bibinfo{author}{\bibfnamefont{D.}~\bibnamefont{Watts}}
  (\bibinfo{collaboration}{Crystal Ball at MAMI and A2 collaboration})
  (\bibinfo{year}{2004}), \bibinfo{note}{11th International Conference on
  Calorimetry in High-Energy Physics, Perugia, Italy.}

\bibitem[{\citenamefont{Audit et~al.}(1991)}]{Audit:1991gq}
\bibinfo{author}{\bibfnamefont{G.}~\bibnamefont{Audit}} \bibnamefont{et~al.},
  \bibinfo{journal}{Nucl. Instrum. Meth.} \textbf{\bibinfo{volume}{A301}},
  \bibinfo{pages}{473} (\bibinfo{year}{1991}).

\bibitem[{\citenamefont{Novotny}(1991)}]{Novotny:1991ht}
\bibinfo{author}{\bibfnamefont{R.}~\bibnamefont{Novotny}}
  (\bibinfo{collaboration}{TAPS}), \bibinfo{journal}{IEEE Trans. Nucl. Sci.}
  \textbf{\bibinfo{volume}{38}}, \bibinfo{pages}{379} (\bibinfo{year}{1991}).

\bibitem[{\citenamefont{Sanderson}(2002)}]{Ruth}
\bibinfo{author}{\bibfnamefont{R.}~\bibnamefont{Sanderson}},
  \bibinfo{journal}{Ph.D. Thesis, The University of Glasgow}
  (\bibinfo{year}{2002}).

\bibitem[{\citenamefont{Tarbert}(2007)}]{Claire}
\bibinfo{author}{\bibfnamefont{C.}~\bibnamefont{Tarbert}},
  \bibinfo{journal}{Ph.D. Thesis, The University of Edinburgh}
  (\bibinfo{year}{2007}).

\bibitem[{\citenamefont{Takaki}(2007)}]{takaki:priv}
\bibinfo{author}{\bibfnamefont{T.}~\bibnamefont{Takaki}},
  \bibinfo{journal}{Private communication}  (\bibinfo{year}{2007}).

\bibitem[{\citenamefont{Watts and MacGregor}(2003)}]{Watts:Cohpi}
\bibinfo{author}{\bibfnamefont{D.~P.} \bibnamefont{Watts}} \bibnamefont{and}
  \bibinfo{author}{\bibfnamefont{I.}~\bibnamefont{MacGregor}},
  \bibinfo{journal}{MAMI proposal Nr A2/5-03}  (\bibinfo{year}{2003}).

\bibitem[{\citenamefont{Krusche et~al.}(2002)}]{Krusche:2002iq}
\bibinfo{author}{\bibfnamefont{B.}~\bibnamefont{Krusche}} \bibnamefont{et~al.},
  \bibinfo{journal}{Phys. Lett.} \textbf{\bibinfo{volume}{B526}},
  \bibinfo{pages}{287} (\bibinfo{year}{2002}).

\bibitem[{\citenamefont{Krusche}(2005)}]{Krusche:2005jx}
\bibinfo{author}{\bibfnamefont{B.}~\bibnamefont{Krusche}},
  \bibinfo{journal}{Eur. Phys. J.} \textbf{\bibinfo{volume}{A26}},
  \bibinfo{pages}{7} (\bibinfo{year}{2005}).

\end{thebibliography}
\end{document}